# A 0.3-2.6 TOPS/W Precision-Scalable Processor for Real-Time Large-Scale ConvNets

Bert Moons, Marian Verhelst
Department of Electrical Engineering, ESAT-MICAS - KU Leuven, Leuven, Belgium
bert.moons@esat.kuleuven.be, marian.verhelst@esat.kuleuven.be

## Abstract

A low-power precision-scalable processor for ConvNets or convolutional neural networks (CNN) is implemented in a 40nm technology. Its 256 parallel processing units achieve a peak 102GOPS running at 204MHz. To minimize energy consumption while maintaining throughput, this works is the first to both exploit the sparsity of convolutions and to implement dynamic precision-scalability enabling supply- and energy scaling. The processor is fully C-programmable, consumes 25-288mW at 204 MHz and scales efficiency from 0.3-2.6 real TOPS/W. This system hereby outperforms the state-of-the-art up to 3.9x in energy efficiency.

## Introduction

Recently CNN's (Fig. 1) have come up as state-of-the-art classification algorithms, achieving near-human performance in speech-recognition and visual-detection [1-3]. However, they are typically very expensive in terms of energy consumption. In [4], an algorithm-level study, we demonstrated opportunities for drastic energy reductions in CNN's through dynamic word length scaling and sparse guarding. Precision requirements vary across CNN's and even across CNN-layers, as the necessary number of bits can go down from 16 to 5 or even 1 bit for different benchmarks, with less than 1% accuracy loss (Tab. 1). Origami [5], Nvidia Tegra [6], and Eyeriss [7] offer non-optimal embedded solutions, as they keep computational precision constant and do not adapt to varying requirements. This work is the first to exploit these opportunities and to implement them in a state-of-the-art CNN architecture. It optimizes energy consumption for any CNN with any precision requirement up to 16-bit fixed point, without sacrificing flexibility, programmability, accuracy or throughput. We hereby empower low power, high-performance embedded applications of computer vision.

## Low Power CNN Processor Design

This CNN-processor achieves scalable low power operation through three key innovations: **(A)** a 2D-SIMD MAC-array with shifted inputs, **(B)** dynamic precision and voltage-scaling and **(C)** guarded data-fetches and -operations. Figure 2 shows the high level processor-overview. It contains a precision-scalable 2D-SIMD array in a voltage-scalable power domain, a total of 148kB on-chip data-, guard- and program memory, max-pool and Rectified Linear Unit vector-arithmetic and a DMA with Huffman compression, all in a fixed power domain. The processor has a custom VLIW and SIMD instruction set and is fully programmable in C using dedicated libraries and a custom generated compiler. The chip is clock-gated and operator guarded where possible to save dynamic power.

*A. The 16x16 2D-SIMD MAC array* (Fig. 3) generates 256 intermediate outputs per cycle while consuming only 16+16 inputs. These MACs are single cycle and contain a 48-bit accumulation register. In an 11x11 convolution example, the MAC-array takes in 16 subsequent pixels from a single image channel and 16 filter weights from different filters in the first cycle. In the next 10 cycles, 17 words are fetched: 16 filter weights and a single pixel, which is shifted through a shift-register. This sequence is repeated once for every row in the 11x11 kernel. Custom instructions allow parallel convolution-execution and data-fetching while all intermediate values are stored in the local accumulation registers. This 2D approach requires 16x fewer data fetches than 1D-SIMD.

An on-chip memory optimized for data-locality (Fig. 4) consists of 64x2kB single-port SRAM macros, subdivided into 4 blocks of 16 parallel banks. 3 blocks can be alternately read or written in parallel: 2 by the processor, a 3rd by the DMA.

*B. Dynamic precision and voltage scaling* can be performed without sacrificing significant accuracy [4]. At lower precision, not only the switching activity drops, but the critical path can become shorter as well if enforced at design-time through a multi-mode optimization (Fig. 3). Positive slack can then be compensated for through a lower supply voltage $V_{scalable}$ while keeping frequency constant. The full 2D-array is placed in a single dynamically scalable power domain. All non-scalable other arithmetic, memories, control and data-transfer are at a fixed supply voltage. E.g. in layer 2 (l2) of the well-established AlexNet benchmark [1], images and filters need 7 bits. This leads to a 1.9x gain compared to full precision and 1.3x additional savings when scaling the supply voltage (Fig. 6).

*C. Guarding memory fetches and operations* allows further energy reductions in sparse CNN computations. The amount of zero-valued data can go up to 89% (Tab. 1). This sparsity is caused both by ReLu operators and by using low precision words [4]. As zero-valued weights or pixels do not contribute to the CNN output, all computations using these values can be skipped. This is done by preventing memory fetches and MAC-operations (Fig. 3). As all sparsity info is known at the start of a new layer, it is stored in a dedicated guard flag memory (Fig. 2, 4). Only 16+16 1-bit flags are fetched to potentially prevent 256 MACs and 32-SRAMs from switching. In AlexNet l2, 19% and 89% zeroes in filters and images lead to an additional 1.9x energy gain (Fig. 6). The Huffman IO-compression (Fig. 2, Tab. 1) reduces bandwidth up to 5.8x for image data and up to 2x overall.

## Measurement Results

This processor was fabricated in a 40nm LP CMOS technology. At room temperature, it runs at a nominal frequency of 204 MHz at 1.1V (Fig.7). It has a total active area of 1.2x2 =2.4mm$^2$. When scaling down from 16- to 8- or 4-bit, the supply voltage $V_{scalable}$ can go down from 1.1V nominally at 204 MHz to 0.9V and 0.8V at 8-bit and 4-bit respectively at the same speed (Fig. 5). All modes run the same non-guarded program, with a typical MAC-efficiency of 77%, achieving up to 2.6 TOPS/W in its most efficient 4-bit mode at 12 MHz.

Three benchmarks: AlexNet [1], LeNet-5 [3] and a general non-scaled/non-guarded 16-bit large scale CNN are run on the processor with measurement results summarized in Table 1. This table shows the precision and guarding opportunities per layer, and the resulting processor memory bandwidth and power reduction. At its nominal frequency the processor dissipates 33mW or 1.6 real TOPS/W for 13400 fps LeNet-5 (28x28 MNIST images) and 76mW or 0.94 real TOPS/W for 47fps AlexNet (227x227 ImageNet images), effectively minimizing energy consumption for both benchmarks. The chip hereby outperforms the non-scalable state-of-the art up to 3.9x in terms of energy efficiency (Fig. 8).

**Acknowledgements**

This work was partly funded by IWT. Special thanks to S. Redant, L. Folens and E. Wouters (imec IC-link) for back-end support.

**References**

[1] A. Krizhevsky, et al., "ImageNet classification with deep Convolutional Neural Networks", *Adv. in NIPS*, 2012.
[2] O. Abdel-Hammid, et al., "Applying CNN concepts to hybrid NN-HMM model for speech recognition", *ICASSP*, 2012.
[3] Y. LeCun, et al., "Gradient-based learning applied to document recognition", *Proceedings of the IEEE*, 1998.
[4] B. Moons, et al., "Energy-efficient ConvNets through approximate computing", *WACV*, 2016. *In press.*
[5] L. Cavigelli, et al., "Origami: a convolutional network accelerator", *Great Lakes Symposium on VLSI*, 2015.
[6] Nvidia Tegra K1 GPU.
[7] Yu- Hsin Chen, et al., "Eyeriss: an energy-efficient reconfigurable accelerator for deep Convolutional Neural Networks", *ISSCC*, 2016. *In press, with author approval.*

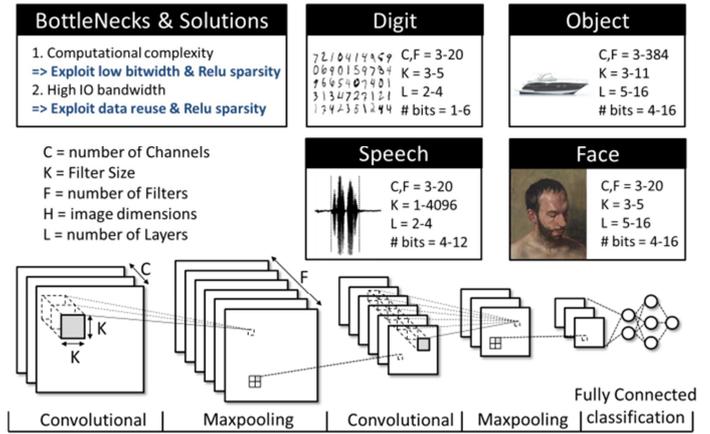

*Fig. 1: **Scalability** in deep convolutional neural networks.*

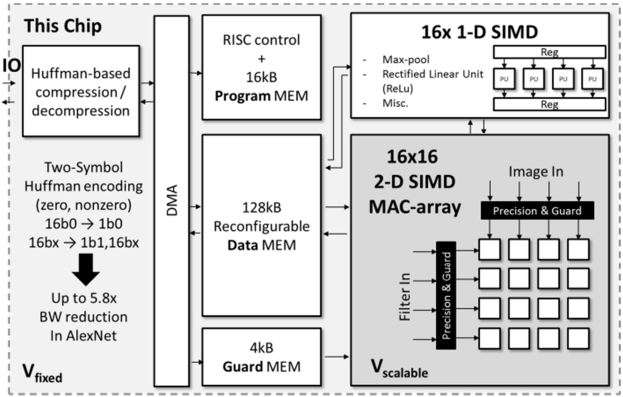

*Fig.2: **Top level architecture**. All non-scalable logic is in a fixed power domain. The MAC-array is in a scalable power domain.*

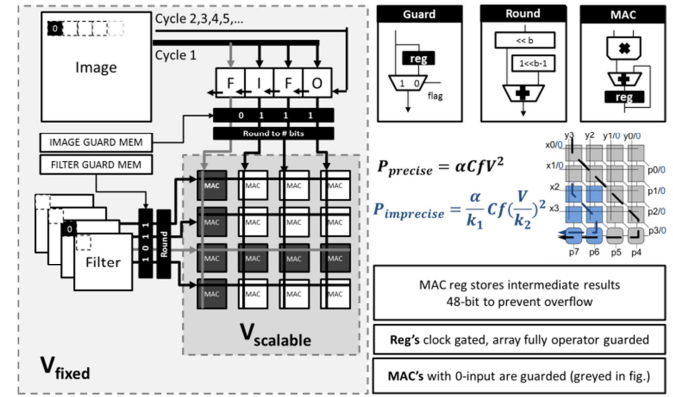

*Fig.3: **2D-SIMD array**. The switching activity and critical path scale with precision. The latter allows lower supply at constant frequency.*

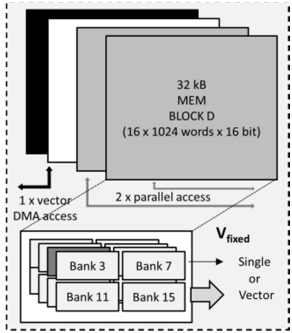

*Fig.4: **Memory architecture**.*

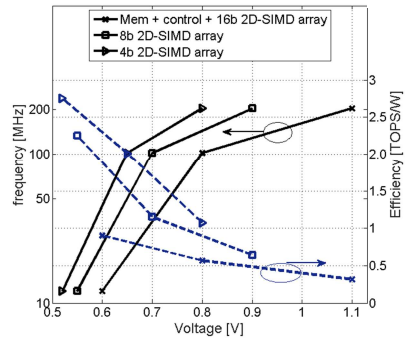

*Fig.5: **Processor performance**. Voltage is for the MAC-array only in the 8b/4b case. Efficiency is for the whole chip.*

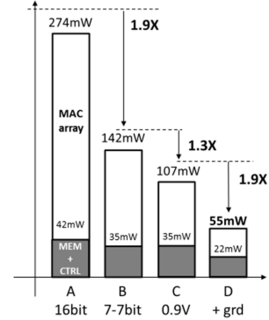

*Fig.6: **Energy saving mechanisms** in AlexNet layer 2. (B) 7-bit (filters) to 7-bit (images) are sufficient. (C) Supply voltage scaling. (D) Guarding added.*

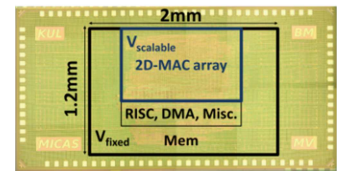

*Fig.7: **Chip photograph**. The processor has an active area of $2.4mm^2$ in 40nm CMOS.*

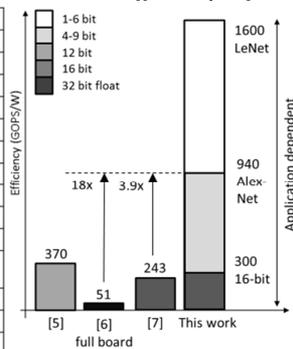

| Technology | 40nm LP (1P_8M) |
|---|---|
| Core Area | 1.2mm x 2mm |
| On-Chip MEM Size | 144kB |
| # MAC's | 256 |
| # Gates (NAND-2) | 1600k |
| Supply voltage | 0.55-1.1 V |
| Leakage | 0.7 mW |
| Frequency | 12-204 MHz |
| Word bit width | 1-16 bit fixed |
| # Filters | All-programmable |
| # Channels | All-programmable |
| Stride | Horizontal: 1-4 Vertical: no limit |
| Peak performance | 102 GOPS |
| Power (AlexNet) | 76 mW |
| Throughput (AlexNet) | 227x227 @ 47fps |

*Fig.8: **Chip overview and comparison**. This work outperforms the state-of-the-art up to 3.9x. Mentioned throughput is for the CNN convolutional layers. [6] is a full Tegra board.[5], [7] and this work are cores only.*

| Layer | Filter / Image bits (0%) | Filter / Image BW Reduc. | IO / HuffIO (MB/frame) | Voltage (V) | MMACs/ Frame | Power (mW) | Real (TOPS/W) |
|---|---|---|---|---|---|---|---|
| General CNN | 16 (0%) / 16 (0%) | 1.0x | — | 1.1 | — | 288 | 0.3 |
| AlexNet l1 | 7 (21%) / 4 (29%) | 1.17x / 1.3x | 1 / 0.77 | 0.85 | 105 | 85 | 0.96 |
| AlexNet l2 | 7 (19%) / 7 (89%) | 1.15x / **5.8x** | 3.2 / 1.1 | 0.9 | 224 | 55 | 1.4 |
| AlexNet l3 | 8 (11%) / 9 (82%) | 1.05x / 4.1x | 6.5 / 2.8 | 0.92 | 150 | 77 | 0.7 |
| AlexNet l4 | 9 (04%) / 8 (72%) | 1.00x / 2.9x | 5.4 / 3.2 | 0.92 | 112 | 95 | 0.56 |
| AlexNet l5 | 9 (04%) / 8 (72%) | 1.00x / 2.9x | 3.7 / 2.1 | 0.92 | 75 | 95 | 0.56 |
| Total / avg. | — | — | 19.8 / **10** | — | — | 76 | **0.94** |
| LeNet-5 l1 | 3 (35%) / 1 (87%) | 1.40x / 5.2x | 0.003 / 0.001 | 0.7 | 0.3 | 25 | 1.07 |
| LeNet-5 l2 | 4 (26%) / 6 (55%) | 1.25x / 1.9x | 0.050 / 0.042 | 0.8 | 1.6 | 35 | 1.75 |
| Total / avg. | — | — | 0.053 / **0.043** | — | — | 33 | **1.6** |

*Tab. 1: **Performance overview** of 16-bit, AlexNet and LeNet-5 CNN's. In AlexNet and LeNet, the used number of bits in filter- and image weights can reduce drastically down to 1-bit, with less than 1% accuracy loss. This data will have high percentages of zero-values (0%), leading to IO BW reduction and more gains through guarding. Supply voltage can be modulated. Not every layer needs an equal amount of MAC-operations/frame. AlexNet l2 needs most operations, leading to a weighted average power consumption of 76mW.*